\documentclass[aps,twocolumn]{revtex4}
\usepackage{graphicx,subfigure}
\usepackage{amsmath,amssymb}
\usepackage{bm}

\newcommand{\be}{\begin{eqnarray}}
\newcommand{\ee}{\end{eqnarray}}

\def\slashchar#1{\setbox0=\hbox{$#1$}           % set a box for #1 
   \dimen0=\wd0                                 % and get its size
   \setbox1=\hbox{/} \dimen1=\wd1               % get size of /
  \ifdim\dimen0>\dimen1                        % #1 is bigger
 \rlap{\hbox to \dimen0{\hfil/\hfil}}      % so center / in box
  #1                                        % and print #1
 \else                                        % / is bigger
    \rlap{\hbox to \dimen1{\hfil$#1$\hfil}}   % so center #1
    /                                         % and print /
 \fi}                                         %

\begin{document}

\title{Comments on  the temperature dependence of the gauge topology}

\author{  Edward  Shuryak }

\affiliation{Department of Physics and Astronomy, Stony Brook University,
Stony Brook NY 11794-3800, USA}

\begin{abstract}
Recent efforts in lattice evaluation of the topological susceptibility had shown that
at high temperatures it is given by well-separated instantons (even in QCD with light fermions,
where those are highly suppressed). Recent development of the semiclassical theory suggest that
below $T_{max}\sim 2.5T_c$, where Polyakov line has values between one and zero, the topology 
ensemble can be represented by a plasma of instanton constituents (called instanton-dyons
or instanton-monopoles). It has been shown that such ensemble undergoes deconfinement and chiral
transitions, semi-qualitatively reproducing the lattice results. There are ongoing efforts to
locate them on the lattice, or use (flavor-dependent) periodicity phases of the deformed versions of QCD on the lattice and
semiclassically,
in order to test this theory.
We here propose another possibly useful tool: the topological susceptibility of a sub-lattice.     
\end{abstract}
\maketitle

\section{Introduction}
 Recently the field of lattice gauge topology has been re-activated, due to two independent
 developments.
 
\begin{figure}[htbp]
\begin{center}
\includegraphics[width=7cm]{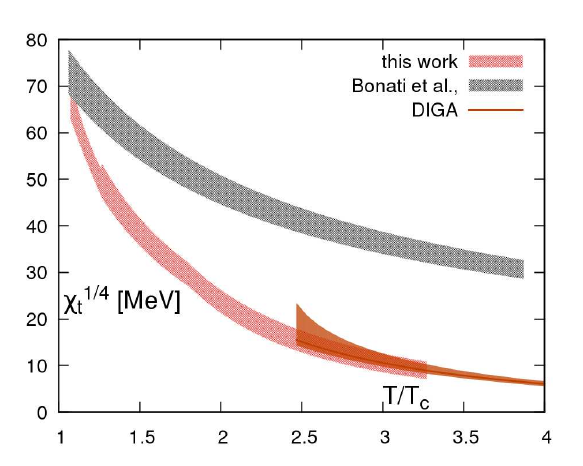}
\includegraphics[width=8cm]{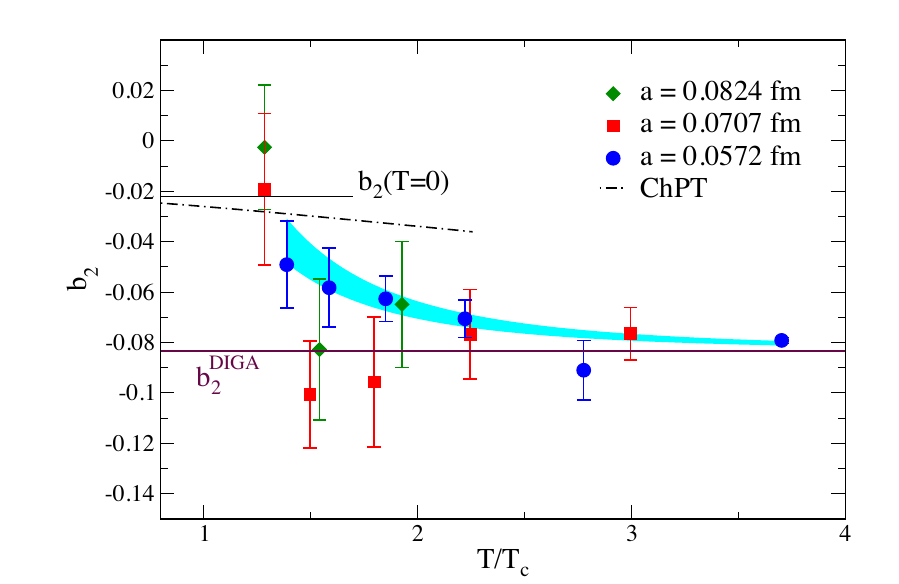}
\caption{(Upper) The topological susceptibility  $\chi^{1/4}$  versus the temperature $T/T_c$. , from \cite{Petreczky:2016vrs}
(Lower) The higher order fluctuation parameter $b_2$, from \cite{Bonati}.}
\label{fig_chi_of_T}
\end{center}
\end{figure}

  One is several recent extensive lattice studies of the topological susceptibility
 $\chi(T)$ in a wide range of temperatures $T$, from zero to  about 2 $GeV$. 
 (Its motivation is partly a relation to axion models of the dark matter.) 
  In Fig.\ref{fig_chi_of_T}(upper)  from \cite{Petreczky:2016vrs} one can see the continuum extrapolated 
  value of $\chi^{1/4}$ versus $T/T_c$ (the lower red shaded region) compared for $T/T_c> 2.5$ with
  the dilute instanton gas approximation (DIGA). The upper gray region corresponds to the results of 
  ref. \cite{Bonati}. Not shown in this plot are results from the work \cite{Borsanyi}, which impressively 
  followed $\chi(T)$ to $T$ as large as $2\, GeV$, also with the conclusion that DIGA is correct at high enough $T$.
  Ref.\cite{Bonati} also had measured higher moment of the topological charge fluctuations, $b_2$.
  The data from this work shown in the lower part of Fig.\ref{fig_chi_of_T} also show that
  for $T/T_c> 2.5$ the DIGA value -- following from $E_{vacuum} \sim cos(\theta)$ -- seem to be reached.
   Accepting these conclusions, we discuss below $why$ the behavior changes below this temperature,
   and what is the correct description of 
   the topology below it.
  
 Another  recent development are works devoted to of the ensemble of the instanton 
 constituents, called instanton-monopoles or instanton-dyons. As shown in the pioneering
 papers by Kraan and van Baal \cite{Kraan:1998sn} and Lee and Lu {Lee:1998bb}, an instanton consists of $N_c$ (number of colors) of those. They share
 unit topological charge of the instanton according to certain fractions $\nu_i, i=1..N_c$ such
 that $\sum_{i=1}^{N_c} \nu_i =1 $. 
 
 The main rational for the instanton to get dis-assembled into those constituents
 is the fact that the mean Polyakov line below certain $T$ deviates from 1, forcing all
 objects to interpolate to a nonzero `holonomy values' of $A_0$. In QCD with physical quarks
 this happens at $T_{Polyakov}\approx 2.5 T_c$ \cite{P_QCD}. 
 
 We suggest that
 deviation of the DIGA from susceptibilities below this temperature  is not a mere coincidence,
 and that at $T< 2.5 T_c$   the instantons are {\em dis-assembled into instanton-dyons}.
  In these comments we discuss how different versions of the
 topological susceptibility can help us to tell if this is indeed the case.

%%%%%%%%%%%%%%%%%%% 
 \section{Various definitions of the topological susceptibility}
 
 In order to see how those theoretical ideas can match lattice measurements,
one needs first to clarify and distinguish various existing definition of $\chi$. 
As we will see, few existing definitions are not at all identical, and should lead to very different results.

We start with the {\em canonical}  definition of the topological susceptibility $\chi_{canonical}$, 
following standard rout of the statistical mechanics. The
 vacuum or thermal ensemble with nonzero theta-angle $\theta$ is defined by 
 an additional term in action, which adds to
the partition function an extra factor $e^{i\theta Q}$, where $Q$ 
is the total topological charge of the volume $V_4$ under consideration. Since $i\theta$ play the same role as
a chemical potential, the topological susceptibility can thus be given
the same standard  definition as any other  susceptibility, namely
\be \chi_{canonical}=({ 1 \over V_4 Z})\left({\partial^2 \over \partial (i\theta)^2}  Z(\theta) \right))   \ee
The volume $V_4$ in this definition should be the one of  a $subsystem$, 
 of even {\em much larger heat bath}. Grand canonical ensemble with the chemical
potential implies that there is free exchange of particles through the boundaries of the subsystem.
%It is the same condition as for canonical ensemble at fixed temperature, for which an exchange of energy
%is required. 
Large  heat bath ensures that the values of variables like $T$ and $\mu=\i\theta$ are fixed, without any fluctuations.

The standard lattice definition 
 \be \chi_{lat}={<Q^2> \over V_4} \ee
 may look identical to the canonical one given above, but is in fact
  quite different, due to the fact that in this case the system with $V_4$ is the whole lattice. 
  It is topologically a torus with $no$ boundaries, since periodic boundary conditions of the fields are imposed. 
  Both electric $Q_E$ and magnetic $Q_M$ charge of  this volume must be zero, and the topological charge $Q$
  must be integer-valued.
  
Another definition $\chi_{sublat}$ has been  proposed by Verbaaschot and myself \cite{Shuryak:1994rr}.
Since it was done many years ago, let us remind it.
We proposed to 
  cut the lattice  into two subsystems, $a$ and $b$, with subvolumes $V_4^a+ V_4^b=V_4$
and define the corresponding susceptibilities by the same expression above. The simplest arrangement
is to cut by two planes normal to one of the coordinates, producing two ``slices" with  \be V_4^a=L^3 x,  V_4^b=L^3 \bar x, \bar x=L-x \ee
This definition needs some\footnote{There are of course
some technical issues to be resolved to get full definition. For example, if one uses fermionic definition of $Q$ based on 
eigenvalues of sub-Dirac matrix, there are questions whether to include links at the boundary (if it includes the lattice plane) or whether to include
links going through it. 
Some experimentation is clearly needed here.
} extra work, but it has two important advantages over the $ \chi_{lat}$.
One is that now the sub-volumes do have a boundary, and they do not have
a requirement that $Q_E=Q_M=0$. As we will discuss below, quarks and Dirac strings can
``leak" through it. 

\begin{figure}[h]
\begin{center}
\includegraphics[width=8cm]{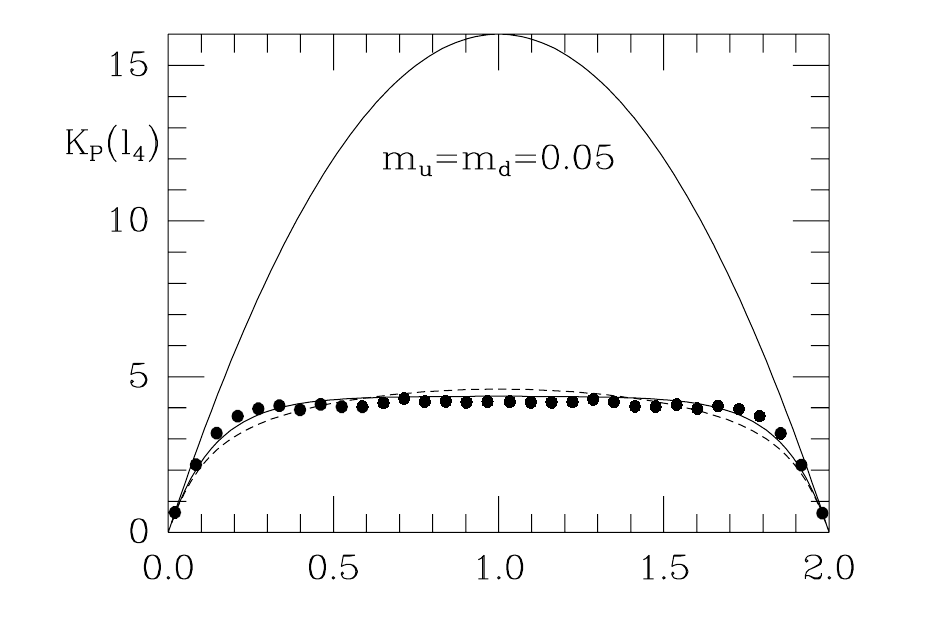}
\caption{An example of the susceptibility in sub-box $\chi_{sublat}$ as a function of the fraction of the total box, $x/L$.
The thin parabolic line corresponds to randomly placed instantons and antiinstantons, the dots are for the interacting instanton liquid, from  \cite{Shuryak:1994rr}. Strong screening of the topological charge in this model is evident.
Thin lines show different fits, from which the value of $m_{\eta'}$ was extracted.
}
\label{fig_screening}
\end{center}
\end{figure}

Note also, that in this setting one obtains not a number but the function  $\chi^a(x)$, which can be used
to define the ``screening length of the topological charge", known also as the $\eta'$ mass.
In this case one gets an idea what is a ``large enough box", since for $m(\eta') x \gg 1$ the dependence on $x$
disappears.

\section{Instantons and the high-$T$ region}

We start discussing the differences between various definitions of $\chi$
 using  the context of the instanton ensemble, in which  $\chi_{sublat}$ was originally introduced.

Let us start with QCD in the chiral limit, with (massless) quarks. In this case any configuration
with nonzero topological charge $Q$ has $Q$ quark zero modes. 
Therefore, the fermionic determinant is zero if $Q\neq 0$,  so the gauge ensemble
include only configurations with $Q=0$ and thus $\chi_{lat}=0$.

Let us first, for simplicity, focus on $T>T_c$, where there is no quark condensate and
the chiral symmetry remains unbroken. In this case the topological objects can exist only as
some clusters with the total  topological charge $Q=0$. The simplest of those are
the instanton-antiinstanton molecules. The ensemble made of those has been
discussed by Ilgenfritz and myself \cite{Ilgenfritz:1988dh}. We do not discuss them here as they are not relevant for 
topological susceptibility.

\begin{figure}[htbp]
\begin{center}
\includegraphics[width=8cm]{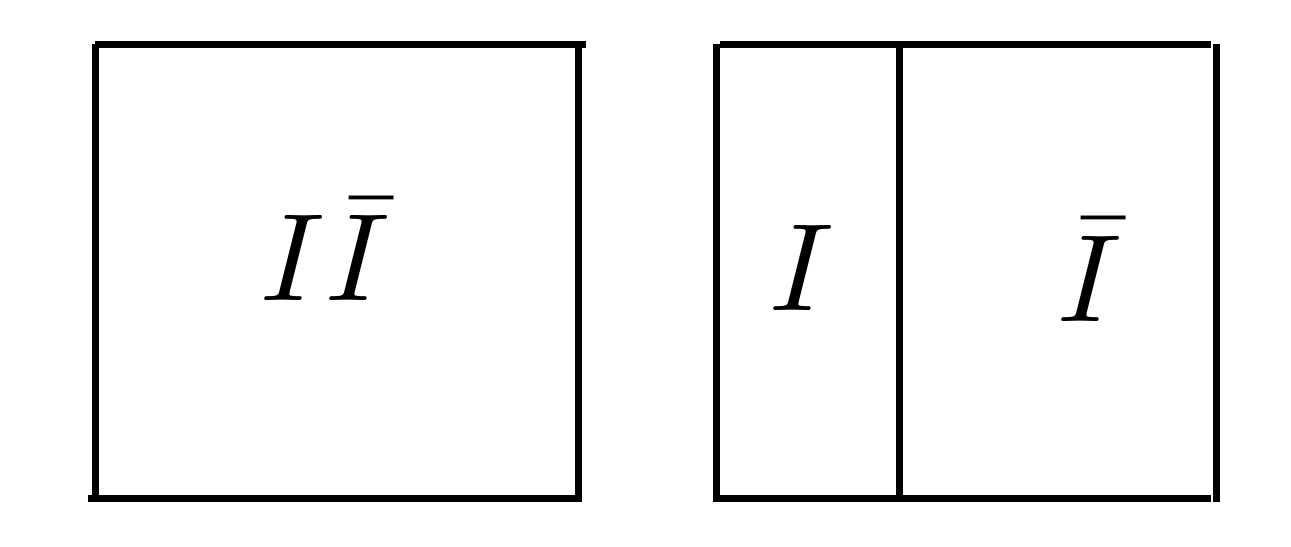}
\caption{Some configurations which produce no contributions  to the $\chi_{lat}$, but contribute to  $\chi_{sublat}$ }
\label{fig_boxes_I}
\end{center}
\end{figure}

While  $\chi_{lat}=0$, the sublattice  definition would lead to a non-zero value $\chi_{sublat}\neq 0$,
because the instantons and the antiinstantons may happen to be located in different subvolumes, see Fig\ref{fig_boxes_I} .
The quarks, created by $I$ and absorbed by $\bar{I}$ may ``leak" through the boundary!

 Note, that for the particular geometry of sub-box proposed, by moving a plane and changing $x$ one changes the sub-volumes $V_4^a, V_4^b$ but $not$ 
the area of the surface $A_3$ separating them. Since the leakage is  expected to be proportional
to this area,  $\chi_{sublat}\sim A_3$, not volume, it should become $x$-independent at large $x$. 

Suppose now we allow small but non-zero quark masses $m$ (for simplicity, the same for $N_f$ quark flavors).
Quark ``veto" on $Q\neq 0$ configurations  such as individual instantons is now lifted.  Since in a dilute ensemble the instantons
can be considered to be non-interacting, one should use the Poisson distribution, and therefore 
\be {\chi(T) \over T^4}\sim  {n(T) \over T^4}\sim ({\Lambda \over T})^b \prod_1^{N_f} ({m_f \over T}) \ee
where $b=11Nc/3+2N_f/3$.
So, the high-$T$ limit corresponds to very small $\chi_{lat}$, decreasing as relatively large power of $T$, times
a rather small product of quark masses.
(We will discuss SU(2) gauge theory, SU(3) gauge theory and QCD with 3 dynamical quarks:
the (inverse) powers of the temperature in those cases are 22/3=7.66, 11 and 6, respectively.)

%This is what is indeed observed in refs  at high $T$.

\begin{figure}[htbp]
\begin{center}
\includegraphics[width=8cm]{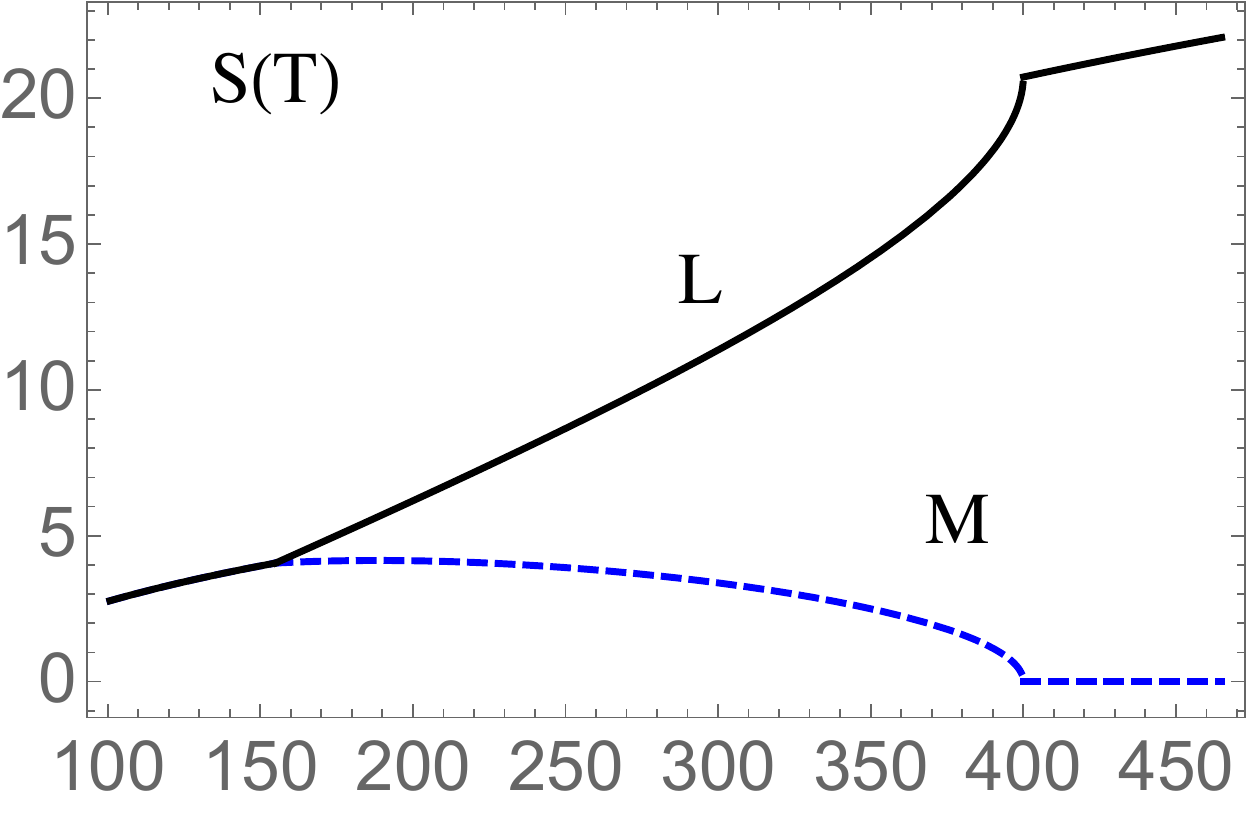}
\caption{The temperature (MeV) dependence of the actions $S/\hbar$ of $L$ and $M$ type dyons, in $SU(3)$ QCD.}
\label{fig_actions}
\end{center}
\end{figure}

\section{The instanton-dyons}
Semiclassical theory of instantons, incorporating a nonzero Polyakov line VEV and thus a nonzero 
 mean value of the gauge field $<A_4>\neq 0$, lead in 1998 to the discovery of the instanton-dyons \cite{Kraan:1998sn,Lee:1998bb}. It is nearly two decades since these papers, but only recently a heavy work on building a semiclassical theory of
their ensemble was intensified. Last year alone has produced about a dozen papers on that. Those will not be
discussed below, for a brief review of some of them see \cite{Shuryak:2016vow}.

When the mean Polyakov line $<P>$ is between 1 to 0, 
 gauge topology is expected to be described by an ensemble
of the instanton-dyons, with different temperature-dependent actions and non-integer\footnote{ This does not violate the usual topological field classification because those require zero field at infinity. The instanton-dyons
 have nonzero magnetic charges and therefore must   be connected by the Dirac strings. 
 Since the Dirac strings are gauge artifacts, ``invisible" for any physical gauge-invariant observable,  they are    
allowed to pass through sublattice boundary. } topological charges,
driven by  $<P(T)>$. 

\begin{figure}[b]
\begin{center}
\includegraphics[width=8cm]{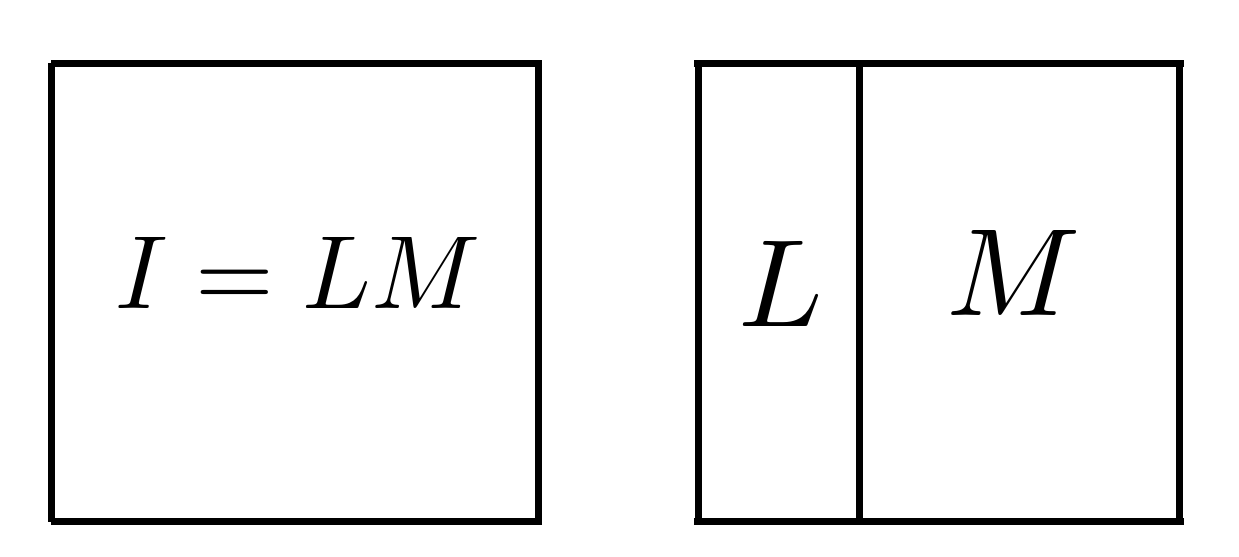}
\caption{Some configurations which produce very different contributions  to the $\chi_{lat}$ and  $\chi_{sublat}$ }
\label{fig_LM}
\end{center}
\end{figure}

In the simplest case of the $SU(2)$ gauge theory there are two types of dyons, selfdual $M$ with $S_M/S_0=\nu $ and $L$ with $S_L/S_0=\bar \nu $, plus anti-selfdual anti-dyons. The parameter $\nu$ is related to the Polyakov line by $<P>=cos(\pi \nu)$. In the $SU(2)$ gauge theory there are two $M$-type dyons, related to complex-conjugated eigenvalues of $<P>$. In this case $\bar \nu =1 -2\nu$ and  $<P>=(1/3)+(2/3)cos(2\pi \nu)$. 

Plugging in the lattice input $<P(T)>$ one can plot the dyon action: see example in Fig.\ref{fig_actions} (for QCD).
This plot can be used to identify the region in which the instanton-dyon theory is semiclassical.
The semiclassical density depends on the action by $n\sim S^2 exp(-S)$: the power of the action represent
half of bosonic zero modes. This formula has a maximum at $S=2$, and we take it as the lowest
possible action at which it makes sense. The left side of Fig.\ref{fig_actions} indicate the lowest temperature 
at which this condition is fulfilled $T_{min}\sim 80\, MeV$. The right side -- high $T$ -- shows that while the
action of the $L$ dyon grows, that of $M$ decreases, so  $T_{max}\sim 370\, MeV$. The phase transition
is indicated as a transition from a symmetric to asymmetric phase. 
These considerations of course refer to simple non-interacting dyons. We use them simply to convey the range of $T_{min}< T <  T_{max}$
in  which this approach is expected to work.

The semiclassical formulae for the 
density of instanton-dyons are higher than instantons, because they have smaller actions. This is the generic reason
why the $\chi(T)$ at $T<2.5T_c$ gets $larger$ than the DIGA prediction. Another generic reason is that $M$-type
dyons have no quark zero modes and thus are not suppressed by fermion masses.

\begin{figure}[h]
\begin{center}
\includegraphics[width=8cm]{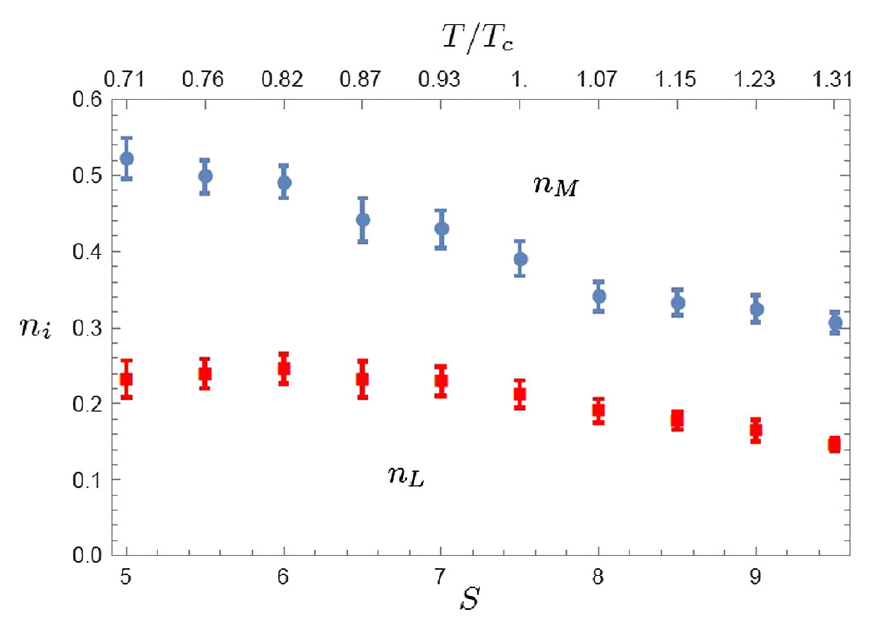}
\includegraphics[width=8cm]{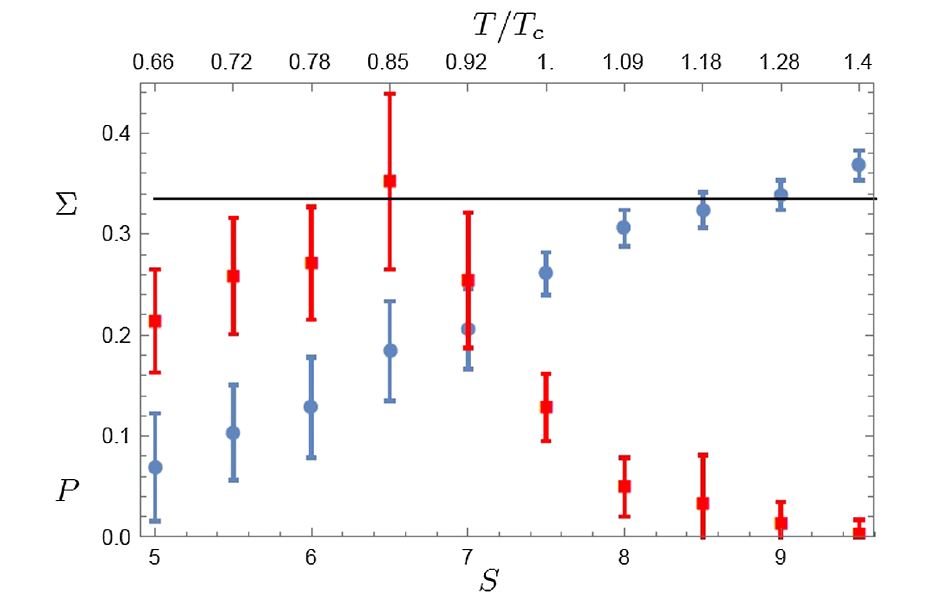}
\caption{(Upper plot)
The densities of  $L$ and $M$ type dyons, versus the action $S=8\pi^2/g^2(T)$, in $SU(2)$ QCD with two light quark flavors. (Lower plot) The mean Polyakov line $<P>$ and quark condensate $\Sigma$, versus the same variable $S$. 
}
\label{fig_densities}
\end{center}
\end{figure}

Proper studies of the dyon ensemble --
such as \cite{Larsen:2015tso} from which we borrowed Fig.\ref{fig_densities} -- include their mutual interaction as well as back reaction to the holonomy potential, determining its value $\nu$
from  the global minimum of the free energy. As one can see from this figure, there is no symmetric phase,
and there are always more $M$ dyons than $L$. Also the deconfinement and chiral transitions become
in QCD-like theories just a smooth cross-overs, happening at roughly the same temperature.

Now let us go back to the topological susceptibilities. Suppose first there are no fermions in the theory. 
Can sub-lattices have non-integer topological charges? Yes:  the Dirac string can penetrate through the boundary
and  nothing prevents a configuration shown in Fig\ref{fig_LM}(right). 
So, $\chi_{sublat}$ would obtain contributions with $non-integer$ values of $Q_a,Q_b$.

This is in sharp contrast to $\chi_{lat}$, since  
lattice configurations can only have $integer$ values of $Q$. At this point, one always ask why in the dyon theory
  any lattice configurations have integer $Q$.  It is because the lattice, unlike the sublattice,
   must have {\em zero total magnetic charge} $Q_{magnetic}=0$.
%In particular, the whole lattice cannot contain a $single$ dyon, since its Dirac string have nowhere to go. 
% The lowest allowed topological states are  (in the SU(2))
%the $LM$ or $\bar{L}\bar{M}$ pairs,  with $Q=1$ and $Q=-1$, respectively. 

This simple discussion nicely illustrates the drastic difference between 
the topology on the lattice and the sub-lattices: the former are
not canonical but in some way share properties of the microcanonical ensembles, with fixed charges.

Let us no return to QCD, switching on  the light quarks. The crucial observation  is that only twisted $L$-type dyon has 
physical -- anti-periodic -- quark zero modes. Therefore,  if quarks are massless, 
those can only exist inside ``neutral clusters", such as $\bar{L} L$ molecules.  We however will not discuss the
molecular component here, as any topologically neutral objects are irrelevant for the topological susceptibility.

%The $M$-type dyons are free from fermions. Furthermore, their action goes to zero as $T\rightarrow T_P$. Therefore,
%naively one may think that in the temperature interval under consideration $T_c>T>T_P$ they would proliferate
%so much that $\chi(T), n_M(T)$ would have a peak. Simulations preformed in \cite{}, see Fig. , show that it is not the case.
%The reason is small-action $M$ dyons have large size: their density is regulated by a repulsive core discovered in \cite{}. 
% 
% At $T\approx T_P$ and above, the action of the $M$-dyon gets small and they are no longer are subject to 
%a semiclassical theory. The $L$ dyons basically become full instantons, so that in this region he dyon theory should match to
%that of dilute instanton gas. 
%

\section{The  topological screening and the $\eta'$ mass }
%In the pure gauge theories the confining phase has $<P>=0$,  $\nu=1/N_c$ and all dyons have the same actions and sizes. 
%
%Mean-field theory
 At $T<T_c$ the
chiral symmetry is broken. How exactly it happens from the point of view of topological object has been worked out
in the instanton liquid model, see \cite{Schafer:1996wv} for a review. The nature of the topological objects involved --an  instanton or only its
$L$-type constituent -- is unimportant: any one with a fermionic zero mode is generating the corresponding 't Hooft vertex.
As $T\rightarrow T_c$ quarks travel through 
 longer and longer  chains  of alternating topological objects. The length of a chain scales as $V_4$, not as its
 dimension $(V_4)^{1/4}$, which in the thermodynamical limit become infinite. That is why Dirac eigenvalues reach
 zero and quark condensate is formed. As a result, pions get massless and one can use chiral perturbation theory
 to describe $\chi(T)$ at $T<T_c$. This is all well known and we do not need to describe it.

There are however some issues related to $\chi_{sublet}$ and the  topological screening length $m_{\eta'}$
we would like to comment on. Let us start with the following (well known) puzzle: 
its numerical value $1/m_{\eta'}=1/.958\, GeV\sim 0.2\, fm)$ is several times smaller compared to
the typical distance between the topological objects, e.g. $L$-dyons at $T_{min}$, which is about $1\, fm $.
One may wander if indeed the quark-induced interaction  can generate so strong correlations inside such  chains. 
The calculation for dyons are in progress, and so we can only mention that it indeed worked out in the instanton liquid,
even in its simplest form, see  $\chi_{sublet}$  already shown in Fig.\ref{fig_screening}.

 Another comment refers to 
 the limit of large number
of colors $N_c\rightarrow \infty$. As famously noted by Witten \cite{Witten:1979vv} , in this limit the $\eta'$ is expected to be light,
\be  m_{\eta'}^2 \sim 1/N_c\rightarrow 0 \label{eta'} \ee 
One should also recall that the so called compressibility of the instanton ensembles, the fluctuations of $N(V)=(g^2/32\pi^2) \int d^4x G_{\mu\nu} ^2(x)$, satisfies the following low energy theorem
\be <N(V)^2>-<N(V)>^2={4 \over b} <N(V)> \ee
where $b=11Nc/3+2N_f/3$.
For  $N_c\rightarrow \infty$ the r.h.s. vanishes, which means the quantity $N(V)$ has in this limt no fluctuations. This in tern 
implies, that the isoscalar scalar meson $\sigma$ must become  heavy. 

In the real world QCD with $N_c=3$ these two masses have the opposite relation, \be m_{\eta'}\approx 958\, MeV > m_{\sigma}\approx 500\, MeV\ee 
but  at some $N_c$ they should become equal, and then continue to move, up and down. According to
instanton liquid study by Schaefer,  the $m(\eta')$ does indeed decreases with $N_c$ as in (\ref{eta'}). 
What happens with $m_\sigma(N_c)$ remains unknown. Lattice studies of these issues would be of significant interest.

Witten \cite{Witten:1979vv} and Veneziano \cite{Veneziano:1979ec} famously related the topological susceptibility to the $\eta' $ mass. 
However, their argument is for $\chi$ in the limit of infinite number of colors, not in physical QCD\footnote{So, to use this relation one needs 
to specify the exact relation of units of both theories, which to our knowledge was never clarified. } .   A similar expression for derived in \cite{Shuryak:1994rr} is based on  $\chi_{sublat}(x)$  in {\em physical QCD} and,
ironically,  corresponds to the limit of $small$ (rather then $large$) volume limit.
%this is how it should be, since the relevant scale is $x m_{\eta'}\sim 1$.

\begin{figure}[t!]
\begin{center}
\includegraphics[width=8cm]{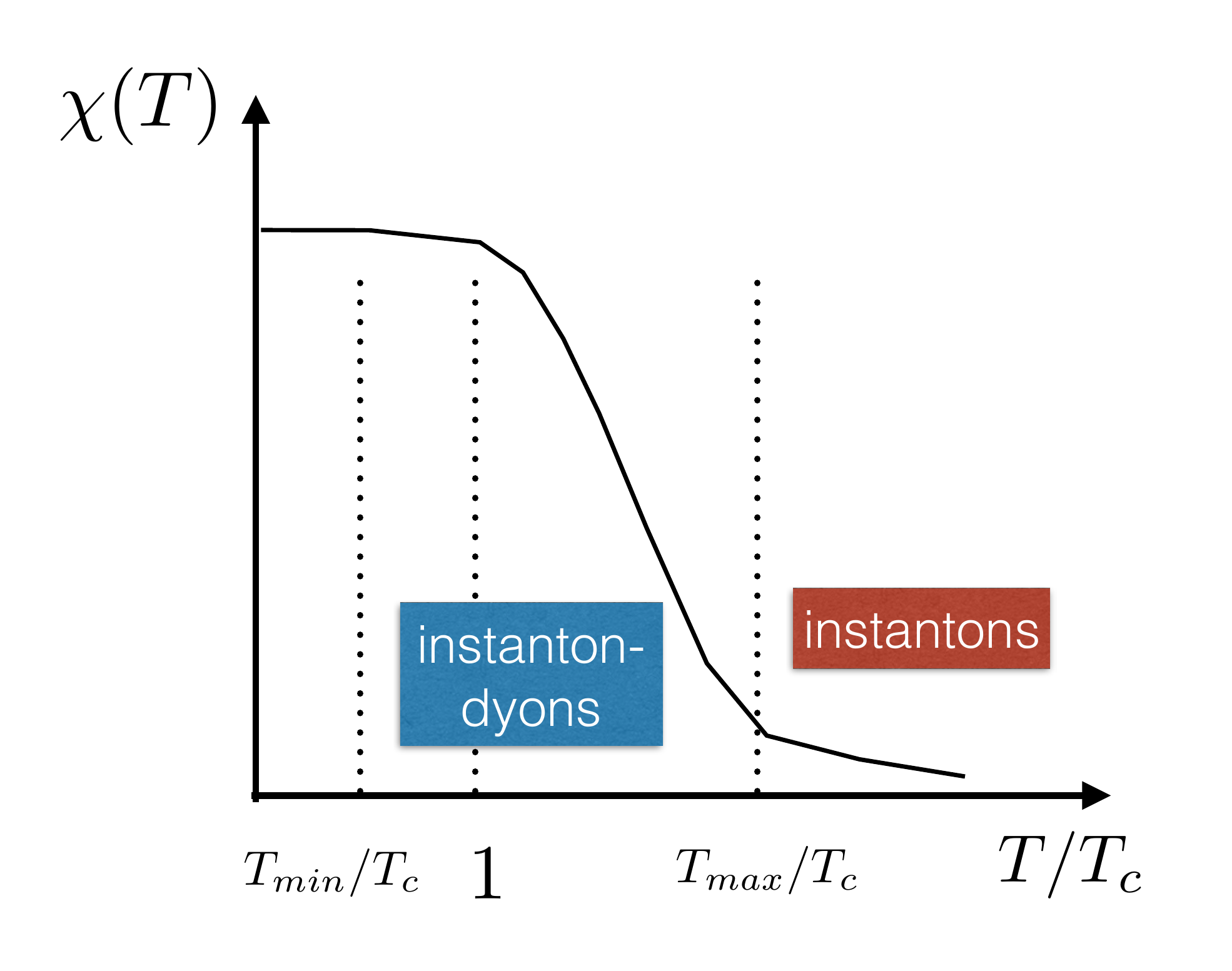}
\caption{A sketch indicating different forms of the topological ensembles as a function of the temperature. 
}
\label{fig_sketch}
\end{center}
\end{figure}

\section{Summary}
These comments can finally be summarized in a sketch shown in Fig.\ref{fig_sketch}:
 below $T\sim T_{max}\sim 2.5T_c$ a dilute instanton gas changes to an ensemble of instanton-dyons.  In this lower region 
$\chi_{lat}\neq \chi_{sublat}$,  they have 
 different $T$-dependences. If it can be evaluated on the lattic,  it will perhaps  reveal the 
 dis-assembly of instantons  into the constituents, with  non-integer topological charges, directly.

\end{document}